\documentclass[twocolumn,prb,superscriptaddress]{revtex4}
\usepackage{graphicx}
\usepackage{dcolumn}
\usepackage{bm}
\usepackage{amsmath}

\begin{document}
\title{Theory of Quantum Optical Control of Single Spin in a Quantum Dot}
\author{Pochung Chen}
\affiliation{Department of Physics, University of California San 
Diego, La Jolla, California
92093-0319}
\affiliation{Department of Chemistry, University of California 
Berkeley, Berkeley, California
94720-1640}
\author{C. Piermarocchi}
\affiliation{Department of Physics, University of California San 
Diego, La Jolla, California
92093-0319}
\affiliation{Department of Physics and Astronomy, Michigan State 
University, East Lansing, Michigan
48824-2320}
\author{L. J. Sham}
\affiliation{Department of Physics, University of California San 
Diego, La Jolla, California
92093-0319}
\author{D. Gammon}
\affiliation{Naval Research Laboratory, Washington, D.C. 20375-5347}
\author{D. G. Steel}
\affiliation{Harrison M. Randall Laboratory of Physics, The 
University of Michigan, Ann Arbor, MI
48109-1120}

\date{\today}

\begin{abstract} We present a theory of quantum optical control of an 
electron spin in a single
semiconductor quantum dot via spin-flip Raman transitions. We show 
how an arbitrary spin rotation
may be achieved by virtual excitation of discrete or continuum trion 
states.  The basic physics
issues of the appropriate adiabatic optical pulses in a static 
magnetic field to perform the single
qubit operation are addressed.
\end{abstract}

\maketitle

\section{Introduction}

Spin-flip Raman spectroscopy has been widely applied to the study of the
properties of donors and acceptors in semiconductors.\cite{yafet73} It was
first used\cite{thomas68} for bound donors in CdS, and
coherent phenomena such as Raman spin-echo were subsequently
observed.\cite{hu76,geschwind84}  Coherent spectroscopic techniques
have attracted new interest due to their potential utilization in the
control and manipulation of simple quantum mechanical systems. In particular,
the application of coherent Raman processes to  qubit operations in
quantum information processing has been suggested for a variety of systems, for
example, an electron spin in a semiconductor quantum dot,\cite{ima} trapped
ions,\cite{plenio,wine} molecules, \cite{lidar}  and rare-earth 
impurities in crystals,
\cite{ham}.
  The optical rotation of electron spins has been demonstrated in 
semiconductor quantum wells.
\cite{awsch,merlin}

In this paper we show how spin-flip Raman optical transitions can 
lead to the full quantum control
of a single electron spin in a semiconductor quantum dot. This 
involves optically connecting the
two electron spin ground states to trions as the intermediate excited 
states. A trion is a bound
state of an exciton with the electron in the dot.  The role of using 
one or more discrete states in
the dot and continuum states in the host are analyzed. The 
constraints in the design of the optical
pulses to preserve the adiabaticity necessary for a high fidelity of 
the control are discussed.  There
are two possible advantages of optical control  compared with other 
control schemes, in
the femtosecond time scale  of theultrafast laser pulses and the 
efficiency and flexibility of pulse
shaping techniques \cite{chen01,pier02} for quantum operations.

The extant experimental situation provides a sound foundation towards 
implementation of our theory.
A semiconductor quantum dot charged with one electron presents a 
strong analogy to a single bound
donor.  However, the spin-flip Raman experiments in semiconductors 
mentioned above involve
ensemble measurements whereas quantum control would require 
experiment on a single dot.
The quantum control of a single exciton in a single dot by coherent 
optical techniques is affirmed
by the experimental demonstration of the Rabi 
oscillations.\cite{stievater01,kamada01,htoon02}
Magneto-luminescence of trion levels from a single dot reported 
recently\cite{tischler02}
forms the basis toward quantum control.

The spin of an electron in a quantum dot has been proposed as a qubit for 
the implementation of
quantum computers.\cite{loss98} It has the advantage of an extremely 
long spin-flip decoherence
time,\cite{gupta99} making it possible to perform a large number of 
quantum operations. We provide
here a full theory for the quantum control of single qubit operations 
using optical pulses.  When
combined with the proposal to couple spins in neighboring quantum 
dots by optically induced RKKY
interaction,\cite{carlo02} one has a complete scheme to build a 
scalabe quantum computer based on
spins in quantum dots via optical control. Although in principle the 
optically controlled RKKY
interaction alone is sufficient for universal computation, 
\cite{divincenzo00} the requirement of
at least three physical qubits to form a single qubit makes the route 
of using a complete set of
single qubit operations plus a two-qubit conditional operation 
perhaps less difficult for the
purpose of a minimalist physical demonstration of two-qubit 
``computation''. The idea of using
Raman schemes to realize single qubit operation was mentioned by 
Imamoglu {\it et
al.}\cite{ima} and Pazy {\it et al.}\cite{pazy03}. Here we expand 
this suggestion, providing
a full theory of single spin rotation by means of optical pulses with 
an explicit formulation.

The paper is organized as follows. In section~\ref{trion} we discuss 
the selection rules and the
effects of an external static magnetic field in optical transitions 
involving trions in quantum
dots (QD). While we focus on QDs generated by mono-layer fluctuations 
in narrow quantum
wells\cite{gammon96} as an example, the theory is applicable to other 
kinds of dots, such as
self-assembled dots. We introduce one particular configuration of 
light polarization and magnetic
field orientation that realizes a lambda system. Section~\ref{raman} 
explains how to perform
adiabatic Raman transitions in this lambda system via one trion 
state. The link between the
parameters of the optical pulses and the angle and the axis of the 
spin rotation is given in the
most general case. The dependence of the spin rotation on the 
orientation of the magnetic field is
presented. The suppression of decoherence in the adiabatic regime is 
shown by a numerical solution
  of the dissipative dynamics based on the Liouville equation and 
explained by a qualitative discussion.
Section~\ref{more} examines the effects of the adiabatic Raman 
transitions via multiple discrete
or continuum trion states. Section~V summarizes the key results.

\section{Trion states in a charged dot}
\label{trion}

We  consider a system of electrons and holes confined in a quantum 
dot described by the Hamiltonian
\begin{equation} H^{eh}=H_0+H_{Coul}+H^e_B+H^h_B+H_C(t),
\end{equation} where $ H_0$ represents the part of the 
non-interacting electron and hole states and
$\ H_{Coul}$ the Coulomb interaction between them.  The effects of 
the external magnetic field
  on the electrons and holes are given by
\begin{eqnarray} H^e_B&=& \frac{1}{2}  \mu_B \sum_{nj\alpha\beta}g_j^e
e^\dagger_{n\alpha} \sigma^j_{\alpha \beta}  e_{n\beta} \nonumber \\
\text{and~~~~~}
  H^h_B&=& \frac{1}{2} \mu_B\sum_{mj\alpha\beta}g^h_j  B_j
h^\dagger_{m\alpha}  \sigma^j_{\alpha \beta} h_{m\beta} ,
\end{eqnarray}
	where $\hbar$ is set to unity, $\sigma^j_{\alpha \beta}$ 
denotes the $\alpha \beta$-th element
of the Pauli matrix in the Cartesian direction $j$ ($=x,y,z$), and 
$e_{n\alpha}$ ($h_{n\alpha}$)
represents the annihilation operator of an electron (hole) in the dot 
at the $n$-th level and spin
(pseduospin) $\sigma$ up or down. Note that the hole levels include 
the doubly degenerate heavy and
light hole states.
Although in some III-V compounds such as GaAs the electron 
$g$-tensor, $g^e$, is approximately
isotropic,\cite{tischler02}  we allow here for the anisotropic case 
with the principal axes along the
Cartesian axes with $z$ being in the growth direction of the 
semiconductor heterostructure.
  In the dipole and rotating wave approximation the light-matter
interaction is
\begin{equation} H_C(t)= \sum_{i \sigma}\Omega_{i\sigma}(t) 
e^{-i\omega_\sigma t}
e_{i\sigma}^\dagger h_{i\sigma}^\dagger +h.c. ,
\end{equation} where $\Omega_{i \sigma}$ denotes a time-dependent 
complex Rabi frequency
following the envelope of the optical pulse centered at the frequency
$\omega_\sigma$,
propagating in the growth direction with circular polarization 
$\sigma$ (left-handed $\sigma = -1$ and
right-handed $\sigma = +1$). For simplicity, in the heavy-hole 
exciton associated with polarization
$\sigma$, the conduction electron spin component is taken to be 
dominated by $e_{n,-\sigma}$ (spin
$-\sigma 1/2$) and the valence hole $h_{m\sigma}$ (spin $\sigma 
3/2$). For the $\sigma$ light-hole
exciton, the components are $e_{n,\sigma}$ and $h_{m\sigma}$ (spin 
$\sigma 1/2$). Correction of
this simplification is straightforward in computation
\cite{yang} but will unnecessarily complicate the exposition of the 
optical processes below.
  The interaction
$H_C(t)$ represents the control Hamiltonian to be designed for the 
manipulation of the spins.  The
semiclassical approximation is appropriate since the intensity of the 
laser field involved is
strong enough to render the photon fluctuation effects negligible. 
The combined effects of the spin-orbit interaction and the dot 
confinement depress the light
hole levels by tens of meV in these nanostructures, allowing us to 
restrict most of our discussions only to
topmost (one or two) heavy hole levels. See Sections~IIIA and IV.

Consider first the minimal model in which there is only one electron 
level and one hole level in
the quantum dot. This is a reasonable assumption  since
the corresponding exciton is well isolated from the higher states. 
In a dot charged with one
electron, there are two ground states $e^\dagger_-|G\rangle$ and 
$e^\dagger_+|G\rangle$ which
represent the spin-up and spin-down  states of the doped electron 
with respect to the $z$ direction.
$|G\rangle$ denotes the ground state of the quantum dot in the 
absence of the electron. There are
two trion states $e^\dagger_- e^\dagger_+ h^\dagger_+|G\rangle$ and
$e^\dagger_+ e^\dagger_- h^\dagger_-|G\rangle$. In the basis of
$e^\dagger_-|G\rangle,e^\dagger_+|G\rangle,e^\dagger_- e^\dagger_+ 
h^\dagger_+|G\rangle$ and
$e^\dagger_+ e^\dagger_- h^\dagger_-|G\rangle$ the Hamiltonian 
including the effect of external
magnetic field and light-matter interaction has the form
\begin{widetext}
\begin{equation} H= \left[\begin{array}{cccc} \omega_B g^e_z \cos\theta &
\omega_B g^e_x \sin\theta & \Omega_+^* e^{i\omega_+ t} & 0 \\ 
\omega_B g^e_x \sin\theta & -\omega_B
g^e_z \cos\theta & 0 & \Omega_-^* e^{i\omega_- t} \\ \Omega_+ 
e^{-i\omega_+ t} & 0 & E_T+\omega_B
g^h_z
\cos\theta & \omega_B g^h_x \sin\theta \\ 0 & \Omega_- e^{-i\omega_- 
t} & \omega_B g^h_x \sin\theta
& E_T-\omega_B g^h_z \cos\theta \\
\end{array}\right],
\label{eq:hammat}
\end{equation}
\end{widetext}
where  $\omega_B= \frac{1}{2}\mu_B |\vec{B}|$ and $\theta$ is the angle between
external magnetic field and $z$-axis.  $E_T$ is the excitation energy 
of the trion state at zero
magnetic field.  The Hamiltonian can be used to calculate the linear 
absorption spectra of trions
in various magnetic field configurations. For the heavy hole, $g^h_x$ 
is negligible if the heavy
hole-light hole mixing and the $k^3$ terms in the Luttinger 
Hamiltonian are neglected.  The
Hamiltonian in Eq.~(\ref{eq:hammat}) is the same one used  by Tischler {\it et
al.}\cite{tischler02} to deduce the g tensors from the 
magneto-photoluminescence measurements.

\begin{figure}
\includegraphics[scale=0.35,angle=270]{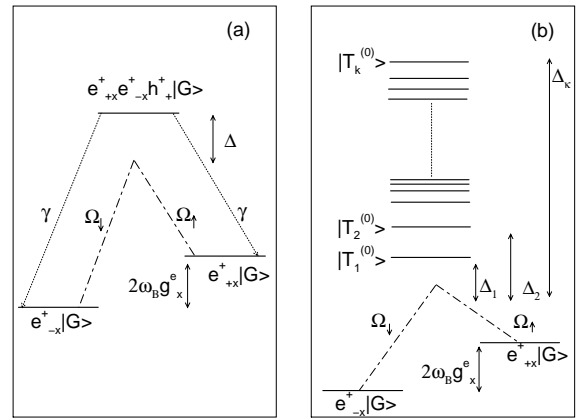}
\caption{Lambda systems in a QD. Only $\sigma+$ polarized light is 
used in the Voigt configuration.
(a) Single trion model. (b) Multiple trion-level model.  At low 
temperatures, the main decoherence
mechanism is the spontaneous radiative decay of the trion state 
indicated by $\gamma$ in (a).
$\Omega_{\uparrow}=\Omega_{1}/\sqrt{2}$, and
$\Omega_{\downarrow}=\Omega_{2}/\sqrt{2}$ are defined in 
Eq.~\ref{phaselock} satisfying the
Raman condition in Eq.~\ref{ramanc}.}
\label{fig1}
\end{figure}

It is clear from Eq.~(\ref{eq:hammat}) that the two spin ground 
states are not coupled by the
applied oscillating electric field unless there is a mixing magnetic 
field tilted away from the
$z$-axis.   We shall present first the simple case of the Voigt configuration
$\theta=\pi/2$ in which the magnetic field is in the quantum well 
plane with its direction
designated as the $x$-axis. Generalization to arbitrary field 
direction (see Section IIIB) is
straightforward. The Voigt case is worth special attention because it 
is the simplest case for
experimental implementation and it gives the simplest illustration of 
the underlying physics for
the control of the single qubit operation. In the case where only
$\sigma_+$ polarized light is used and setting
$g^h_x=0$,  the trion state $e^\dagger_+ e^\dagger_- 
h^\dagger_-|G\rangle$ is decoupled from the
rest.  The magnetic field in the
$x$ direction produces a Zeeman splitting between the states
$e^\dagger_{\pm x}|G\rangle=(1/\sqrt{2})(e^\dagger_- \pm 
e^\dagger_+)|G\rangle$.  The states
$e^\dagger_{+x}|G\rangle$, $e^\dagger_{-x}|G\rangle$ and
$e^\dagger_-e^\dagger_+h^\dagger_+|G\rangle=e^\dagger_{-x}e^\dagger_{+x}h^\dagger_+|G\rangle$
identify a three level system. Consider now two phase-locked 
$\sigma_+$ polarized lasers pulses
which give rise to an off-diagonal matrix element of the interaction 
Hamiltonian in
Eq.~(\ref{eq:hammat}) of the form
\begin{equation}
\Omega_+(t) =\Omega_1(t)e^{i(\omega_+ -\omega_1) t - i
\alpha}+\Omega_2(t)e^{i(\omega_+ - \omega_2) t}~,
\label{phaselock}
\end{equation} where $\alpha$ is the relative phase between the two real Rabi
energies $\Omega_{1}(t),\Omega_{2}(t) $. This form of the pulses can 
be obtained with pulse-shaping
techniques.   The frequencies $\omega_{1},~\omega_{2}$ are chosen to 
satisfy the Raman conditions,
\begin{equation}
\omega_{1}+\omega_B g^e_x=\omega_{2}-\omega_B g^e_x= \omega_+ \equiv 
E_T-\Delta~, \label{ramanc}
\end{equation}  where $\Delta$ is the common Raman detuning (see 
Fig.~\ref{fig1}(a)). In the
rotating frame defined by
$e^{\mp i\omega_B g^e_x t}e^\dagger_{\pm x}|G\rangle$ and
$e^{-i(E_T-\Delta)t}e^\dagger_{-x}e^\dagger_{+x}h^\dagger_+|G\rangle$, 
the Hamiltonian becomes
\begin{widetext}
\begin{equation} H=\frac{1}{\sqrt{2}}\left[\begin{array}{ccc} 0 & 0 &
\Omega_{1}(t)e^{i\alpha}+\Omega_{2}(t)e^{2i g^e_x\omega_B t} \\ 0 & 0 &
\Omega_{1}(t)e^{-2i g^e_x \omega_B t +i\alpha}+\Omega_{2}(t)\\
\Omega_{1}(t)e^{-i\alpha}+\Omega_{2}(t) e^{-2i g^e_x\omega_B t} 
&\Omega_{1}(t)e^{2i g^e_x \omega_B
t -i\alpha}+\Omega_{2}(t)&
\sqrt{2}\Delta
\end{array}\right].
\label{freqsel}
\end{equation}
\end{widetext}  When $|\Omega_j(t)| \ll  g^e_x \omega_B$, the fast 
oscillating terms can be
neglected.  Then the Hamiltonian takes the form
\begin{equation}
\label{H_3level} H_r = \left[\begin{array}{ccc} 0 & 0 & 
e^{i\alpha}\Omega_{\uparrow}(t)  \\ 0 & 0 &
\Omega_{\downarrow}(t) \\ e^{-i\alpha}\Omega_{\uparrow}(t)  & 
\Omega_{\downarrow}(t)  & \Delta \\
\end{array}\right]~,
\end{equation} where $\Omega_{\uparrow}=\Omega_{1}/\sqrt{2},
\Omega_{\downarrow}=\Omega_{2}/\sqrt{2}$.
   This constitutes a single $\Lambda$ system as shown in 
Fig.~\ref{fig1}(a).  For typical Zeeman
splittings of 1 meV and  simple Gaussian pulses, frequency 
selectivity requires a pulse temporal
width much longer than 0.6~ps (= $\hbar/$ 1 meV).

\section{Control of spin dynamics in a charged dot}
\label{raman} Stimulated Raman adiabatic passage 
(STIRAP)\cite{STIRAP} has been extensively used to
perform population transfer between quantum states.  \cite{vitanov97} 
It has also been used to
create entangled states.\cite{unanyan01} In contrast to the typical 
STIRAP population transfer
scheme, we do not make assumptions on the initial state of the 
system. The transformation we are
considering are general rotations, independent of the initial 
orientation of the spin.  A procedure
to perform general spin rotation via STIRAP was proposed 
recently.\cite{kis02} However, an extra
auxiliary ground state was required in addition to the two ground 
states. It is unsuited to the
case of a single charged quantum dot. This STIRAP method can be used 
in coupled QDs for single
qubit operations and quantum gates.\cite{troiani03,calarco03} In the 
following we will first show
how to perform an arbitrary spin rotation in a single $\Lambda$ 
system without using any auxiliary
level. We will discuss then the adiabatic condition and the effect of 
the decaying intermediate
trion state.

\subsection{General single-spin rotation}
\label{raman_3level}

The Hamiltonian of the single $\Lambda$ system in 
Eq.~(\ref{H_3level}) may be diagonalized
analytically by the substitutions,
\begin{eqnarray}
\Omega_\uparrow &=&\Xi \sin(2\phi) \cos\beta, \nonumber \\
\Omega_\downarrow &=&\Xi \sin(2\phi) \sin\beta, \label{eq-sub} \\
\Delta &=& 2 \Xi \cos(2\phi). \nonumber
\end{eqnarray}
$\Xi(t)$ is the grand Rabi frequency,
\begin{equation}
\Xi = \sqrt{\Omega_{\uparrow}^2+\Omega_{\downarrow}^2 
+\left(\frac{\Delta}{2}\right)^2}.
\end{equation}
  The angle $\phi(t)$ may be called roughly the tipping angle of the 
pseudo-magnetic field if the
three states are regarded as pseudo-spin states. To make clear the 
physical meaning of
$\beta$  below, it is convenient  to make the two pulses, 
$\Omega_{\uparrow}(t)$ and
$\Omega_{\downarrow}(t)$, with the same envelope shape. Then the angle $\beta
=\arctan(\Omega_{\downarrow}/\Omega_{\uparrow})$ is independent of 
time. In general,
  the pulse shape identity may be relaxed to the extent that the time 
independence of $\beta$
becomes a slowly varying one to satisfy the  adiabatic condition to 
be considered next. The matrix of
three columns of eigen-vectors,
\begin{equation} W(t) = \left[ \begin{array}{ccc} - e^{i\alpha} 
\sin\beta & - e^{i\alpha} \cos\beta
\cos\phi &  e^{i\alpha}
\cos\beta\sin\phi \\  \cos\beta  & -\sin\beta\cos\phi & 
\sin\beta\sin\phi \\ 0 & \sin\phi &
\cos\phi \\
\end{array} \right],\label{adstates}
\end{equation} leads to the diagonal form of the Hamiltonian at time 
$t$, $W^\dagger(t) H_r(t)
W(t)$, with the eigenvalues along the matrix diagonal, respectively,
\begin{subequations}
\begin{eqnarray}
\lambda_1(t) &= & 0, \label{l1} \\
\lambda_2(t) &=& -2\Xi(t)\sin^2\phi(t), \label{l2} \\
\lambda_3(t) &=& 2\Xi(t)\cos^2\phi(t). \label{l3}
\end{eqnarray}
\end{subequations}
The time-dependent eigenstates are used to form an adiabatic basis set. The
effective Hamiltonian in this representation for the time-dependent 
Schr\"odinger equation is given by
\begin{equation} H_{ad}=W^\dagger H W-iW^\dagger \frac{dW}{dt}. \label{adH}
\end{equation}
If the second term on the right is neglible, the transformed 
Hamiltonian is diagonal, leading to a
diagonal evolution operation  $U_{ad}(t, t')$ with the terms
$e^{-i\Lambda_j}, j=1,2,3$, where $\Lambda_i=\int^t_{t'} dt'' 
\lambda_i(t'')$. Since
$\Lambda_1=0$,  the first eigenstate is  time independent and 
completely decoupled from the other
two states. The motion governed by the instantaneous eigenenergies is 
known as adiabatic. The condition
for the adiabatic approximation is the slow time variation of $W$ 
which, from Eq.~(\ref{adstates}),
depends on the rate of change of the tipping angle
$\dot{\phi}(t)$  in comparing with the rate of the adiabatic motion given by
the grand Rabi frequency which sets the magnitudes of the 
instantaneous eigenenergies,
\begin{equation} |\dot{\phi}(t)|\ll 2\Xi(t).
\label{adiabaticcond}
\end{equation}

  For the qubit operation, at $t=-\infty$ the state of the system is a 
linear combination in the
subspace spanned by the eigenstates associated with
$\lambda_1$ and $\lambda_2(-\infty)$. The time dependent Hamiltonian 
describing the optical pulses
has a cyclic behavior, meaning that  $H(t=\infty)=H(t=-\infty)$. The 
idea of the adiabatic evolution is
that, if the Hamiltonian varies slowly enough in time, the state of 
the system remains
confined in the subspace spanned by the two eigenstates
  at all times. An arbitrary initial state in the spin ground state subspace,
$[a, b]^T$ will acquire only a phase in the $\lambda_2$ component, 
transforming to
$[a, e^{-i\Lambda_2} b]^T$.  The evolution operator in the original 
rotating frame is given by
\begin{eqnarray} U(+\infty, -\infty) &=& W(\infty)U_{ad}W^\dagger(-\infty) 
\nonumber \\ &=& \left[
\begin{array}{cl} e^{-i\Lambda_2/2} U_2 & \begin{array}{lc} \vdots & 
0 \\ \vdots & 0  \end{array}
\\
\multicolumn{2}{c}\dotfill \\
\begin{array}{ccc} 0 & & 0  \end{array} & \begin{array}{lc} \vdots & 
e^{-i\Lambda_3}
\end{array} \\
\end{array}\right],
\end{eqnarray} where
\begin{equation} U_2 =e^{-\frac{i}{2}\Lambda_2 \vec{\sigma} \cdot 
\vec{n}}, \label{eq-rot}
\end{equation} is our final result  for the rotation in  the spin 1/2 
subspace through an angle
$\Lambda_2$ about the unit vector $\vec{n}$ in the polar direction 
given by the declination and
azimuthal angle, $(2\beta, \alpha)$, or
\begin{eqnarray} n_1&=& \cos\alpha \sin(2\beta), \nonumber \\ n_2&=& 
-\sin\alpha \sin(2\beta), \\
n_3&=&\cos(2\beta), \nonumber
\end{eqnarray} where the Cartesian directions (1,2,3) are along the 
unit vectors
$(\hat{z},-\hat{y},\hat{x})$. The polar direction is along the 
magnetic field. The effect of the
spin precession due to the magnetic field is avoided by always 
working in the rotating frame
introduced by Eq. (7).

The corrections due to the light-hole come in two forms. One is the 
light-hole mixing in the heavy-hole
and electron singlet-pair trion state. \cite{yang} The spin-up 
electron is connected
  by the $\sigma_+$ polarization to the $+3/2$ heavy-hole trion whose 
mixture with the $+1/2$
light-hole component is connected by the same polarized  light to the 
spin down electron state. This
induces an extra rotation of the order of 1\% of $\Lambda_2$ about an 
axis normal to the growth
axis, which is just a minor correction which can be included in the 
effect of the transverse magnetic
field. The other correction is due to the light-hole trion whose 
effect is  small if the detuning is less
than 10~meV \cite{calarco03} and can be eliminated by pulse-shaping.
\cite{chen01,pier02}

\subsection{Arbitrary magnetic field orientation} \label{arbitrary}

Since the tilted magnetic field is essential to the complete set of 
single-qubit operations, it is
important to study the dependence of the operation on the field 
orientation. The generalization to
an arbitrary direction follows the same procedure as in Sections II 
and IIIA.  Taking again only
$\sigma+$ polarized light, we need to consider only the two spin 
ground states and one trion state
made out of a spin-up $(+3/2)$ hole and two electrons in a singlet.
We rewrite the reduced Hamiltonian
from Eq.~(\ref{eq:hammat}) in the appearance of a Hamiltonian with an 
effective $g^e$,
\begin{equation} H=\left[\begin{array}{ccc} \omega_B g^e \cos\vartheta &
\omega_B g^e \sin\vartheta & \Omega^*_+(t) e^{i\omega_+t} \\ \omega_B g^e
\sin\vartheta & -\omega_B g^e \cos\vartheta & 0 \\ 
\Omega_+(t)e^{-i\omega_+t} & 0 & E_T+\omega_B
g_z^h \cos\theta \\
\end{array}\right],
\end{equation} where we have set $g_x^h$ to zero and defined the

$\theta$-dependent effective $g^e$
  and the effective angle $\vartheta$ by
\begin{eqnarray} g^e(\theta) &=&\sqrt{(g_z^e\cos\theta)^2+(g_x^e
\sin\theta)^2}, \\
\vartheta(\theta) &=&\arctan\left(\frac{g_x^e}{g_z^e}
\tan\theta\right)~.
\end{eqnarray} By the unitary transformation
\begin{equation} U=\left[\begin{array}{ccc} \cos\frac{\vartheta}{2} &
\sin\frac{\vartheta}{2} & 0 \\ \sin\frac{\vartheta}{2} & 
-\cos\frac{\vartheta}{2} & 0 \\ 0 & 0 & 1
\\
\end{array}\right],
\end{equation} the three basis states are transformed to the spin 
states along the field direction
$|\pm B\rangle$, and the invariant trion $|T\rangle$. When the two 
pulses are chosen as in
Eq.~(\ref{phaselock}), the Hamiltonian
$H_r$ in the new rotating frame is exactly of the same form as 
Eq.~(\ref{H_3level}). The only
changes are in the expressions for the Rabi energies and the detuning,
\begin{subequations}
\begin{eqnarray}
\Omega_\uparrow &=& \Omega_1 \cos\frac{\vartheta}{2}, \\
\Omega_\downarrow &=& \Omega_2 \sin\frac{\vartheta}{2}, \\
\Delta &=& E_T+\omega_B g_z^h \cos\theta- \omega,
\end{eqnarray}
\end{subequations}

The solution then follows exactly the procedure in 
Section~\ref{raman_3level}. The resultant
evolution yields the spin rotation as in Eq.~(\ref{eq-rot}). The Cartesian 
axes (1,2,3) are along
the unit vectors $(-\hat{y}\times\hat{B}, -\hat{y}, \hat{B})$. As a 
check, note that if the
magnetic field is parallel to the propagation axis of the light, then 
$\vartheta=0$ and we can
realize only rotations about the $z$ axis. On the other hand, for a 
finite $\vartheta$ we can
obtain a rotation about any axis by changing the control parameters $\alpha$,
$\Omega_1$, $\Omega_2$ and $\Delta$.

\subsection{Suppression of trion decoherence}

It is physically reasonable that the use of the off-resonance Raman 
processes should avoid the
short optical decoherence time due to the rapid recombination of the 
exciton, since the the excited
state is only virtually excited. The coherence of the spin dynamics 
is then governed by the much
longer spin dephasing time.  In a more quantitative study, we 
consider the eigenstates in
Eq.~(\ref{adstates}). The first eigenstate
$|\lambda_1\rangle$ has no component in the intermediate trion state, 
and the second eigenstate
$|\lambda_2\rangle$ has only a small component in the intermediate 
state as long as
$\Omega(t)/\Delta$ is small. As a result the intermediate state is 
only weakly populated during the
Raman transition and its decoherence has a weak effect on the 
coherence of the spin rotation.

To substantiate this claim, we start with the master equation of the 
density matrix  $\rho$,
\begin{equation}
\frac{d\rho}{dt}=-i[H,\rho] - \frac{1}{2}\sum_i \left( L^\dagger_i L_i
\rho+\rho L^\dagger_i L_i - 2 L_i\rho L^\dagger_i \right) ,
\end{equation} where $L_i$ are the Lindblad\cite{lindblad} operators. 
These operators have the form
of projectors and describe the effect of the spontaneous radiative 
recombination of the trion state
as shown in Fig.~\ref{fig1}(a).  The density matrix in the adiabatic 
representation is
$\varrho=W^\dagger \rho W$ (note the slightly different symbol
$\varrho$ used on the left) and satisfies the transformed equation
\begin{eqnarray}
\frac{d\varrho}{dt}&=& -i[H_{ad},\varrho] \nonumber \\ & & - \frac{1}{2}\sum_i
\left( M^\dagger_i M_i \varrho+\varrho M^\dagger_i M_i - 2 M_i\varrho 
M^\dagger_i \right) ,
\end{eqnarray} where $M_i=W^\dagger L_i W$. The effect of the 
transformation on the Lindblad
operators is considerably simplified if we assume that the 
spontaneous emission rates from the
trion to the two spin ground states are the same, $\gamma$. By 
symmetry the results are independent
of the rotations associated with the relative phase of the two pulses 
$\alpha$ and the rotation of
the spin basis states to the magnetic field direction $\beta$. The 
total relaxation part is given by
\begin{widetext}
\begin{equation} \label{relax} M_{\mbox{relax}}[\varrho]= - \gamma 
\left[\begin{array}{ccc} 0 &
r^\dagger_1 \sin\phi & r_1^\dagger \cos\phi \\ r_1 \sin\phi & (r_2 + 
r_2^\dagger) \sin\phi &
r_2^\dagger
\cos\phi + r_3 \sin\phi
\\ r_1 \cos\phi & r_2 \cos\phi + r_3^\dagger \sin\phi & (r_3 + r_3^\dagger)
\cos\phi \\
\end{array}\right] + \gamma r_0 \left[\begin{array}{ccc} 1 & 0 & 0 \\ 
0 & \cos^2\phi &
-\sin\phi\cos\phi
\\ 0 &  -\sin\phi\cos\phi & \sin^2\phi \\
\end{array}\right],
\end{equation} where
$r_j = \varrho_{2,j} \sin\phi + \varrho_{3,j}\cos\phi$, $r_j^\dagger 
=\varrho_{j,2} \sin\phi +
\varrho_{j,3}\cos\phi$, and
$r_0 = r_2\sin\phi + r_3 \cos\phi~$, for $j$=1,2,3. The origin of the 
decoherence in the Raman
process may be exhibited by a simpler expression of 
$M_{\mbox{relax}}$ which is obtained by
expansion in powers of the small quantity $\phi(t)$,
\begin{equation} M_{\mbox{relax}}[\varrho]=\gamma \left\{
\left[\begin{array}{ccc}
\varrho_{33} & 0 & -\varrho_{13} \\ 0 & \varrho_{33} & -\varrho_{23}
\\ -\varrho_{31} & -\varrho_{32} & -2\varrho_{33} \\
\end{array}\right] -
\left[\begin{array}{ccc} -\varrho_{23}-\varrho_{32} & \varrho_{13} &
\varrho_{12}  \\
  \varrho_{31} & 0 & \varrho_{22}+2\varrho_{33} \\ \varrho_{21} & 
\varrho_{22}+2\varrho_{33} &
\varrho_{23}+\varrho_{32} \\
\end{array}\right] \phi+O(\phi^2) \right \} .
\end{equation}
\end{widetext}  At the start of a qubit operation, the density matrix 
has the form
\begin{equation}
\varrho(-\infty)=\left[\begin{array}{ccc} \varrho_{11} & \varrho_{21} & 0 \\
\varrho_{21} & \varrho_{22} & 0 \\ 0 & 0 & 0 \\
\end{array}\right].
\end{equation} If the adiabatic condition in 
Eq.~(\ref{adiabaticcond}) is satisfied, then
$\varrho_{j3} \forall j$ remain nearly zero (of first order in 
$\phi$) at all times. To first order
in $\phi$,
$M_{\mbox{relax}}[\varrho]$ is proportional only to $\varrho_{j3}$ in 
the subspace spanned by
$|\lambda_1\rangle$ and $|\lambda_2\rangle$. Hence, the relaxation 
terms in this subspace are of
second order in
$\phi$. This demonstrates a suppression of the optical decoherence effects
  within the adiabatic subspace.

A more quantitative measure of the qubit operation is the 
commonly-used fidelity which is an
overlap of the physical operation versus the ideal. We follow the 
averaging over all possible
initial states in the Hilbert space as was done in 
Refs.~\onlinecite{chen01} and
\onlinecite{pier02}.  To compute the fidelity, we have performed a 
numerical simulation on the
adiabatic spin rotation using the quantum trajectory 
method,\cite{carmichael} which is equivalent
to solving the master equation with the relaxation terms in 
Eq.~(\ref{relax}). We take  the common
shape of the pulses to be Gaussian,
$\propto \exp{-(t/\tau)^2}$. The lifetime of the trion due to 
spontaneous emission
(Fig.~\ref{fig1}) is taken to be $60$ ps. Other forms of dephasing, 
such as that induced by the
electron-phonon interaction, are experimentally found to be 
negligible in the fluctuation quantum
dots.\cite{bonadeo} We simulate the operation of a $\pi$ rotation in 
the spin space. For a Rabi
energy
$\Omega_0=1$ meV and a detuning
$\Delta=5$ meV we find an appropriate pulse duration given by 
$\tau=8.74$~ps. The resultant
fidelity of this operation is $F=0.991$. If the detuning is increased 
to $\Delta=10$ meV, we find
that the adiabaticity condition is better satisfied and the operation 
is more robust against
spontaneous emission.  The fidelity in this case increases to 
$F=0.995$. The price is a longer pulse
duration, $\tau=16.74$~ps for the $\pi$ rotation. This demonstrates 
numerically that the
decoherence of the intermediate trion state can be suppressed using 
an adiabatic control.  Once the
effect of spontaneous emission has been reduced, the spin-flip 
decoherence is the remaining
limiting mechanism for the coherence of the qubit. This time has been 
found to be of the order of
hundreds of nanosecond,\cite{gupta99} and, therefore, we can afford 
to use rather long pulses for
the control.

\section{Multiple trion levels}
\label{more}

In this section we consider (i) what happens if more than one 
electron or hole levels are localized
in the dot, and (ii) how to extend the theory from discrete trion 
levels to a continuum. In
mono-layer fluctuation QDs and some cases of self-assembled dots, 
these continuum states are
provided by delocalized excitons in the quantum well. We shall 
confine ourselves to the case of higher
electron-singlet heavy-hole trion levels. There are light-hole 
effects which can be shown to be small as in
Section~IIIA. There are  two-electron spin singlets and triplets. 
None of these are important if the
detuning from the lowest single trion is small.

\subsection{Multiple Lambda system}

\label{raman_4level}

We assume that the  initial state is still restricted to a linear 
combination of the spin ground
states, $e^\dagger_{1-}|G\rangle$ and
$e^\dagger_{1+}|G\rangle$.  In the presence of many electron and hole 
levels in the dot, the effect
of Coulomb interaction is to renormalize the trion energies and the 
oscillator strength of the
optical transitions.  Consider again the Voigt configuration. We have two
$\sigma_+$ laser pulses satisfying the two photon coherence configuration:
$\omega_{1}+\omega_B g^e_x=\omega_{2}-\omega_B 
g^e_x=E_{T,1}-\Delta_1$ where $E_{T,1}$ is the
lowest trion eigenstate energy including the effects of the Coulomb 
interaction.  In this case,
there are many possible trion states
$T_1^{(0)} \cdots T_k^{(0)}$ resulting from the many confined levels, 
and one ends up with a
multiple $\Lambda$ system, as depicted in Fig.~\ref{fig1}(b). The 
general theory developed in
section
\ref{raman} can be extended to treat the multiple
$\Lambda$ system.\cite{vitanov99} To illustrate the method we 
consider the case where the two
pulses are identical, i.e., $\Omega_{1}(t)=\Omega_{2}(t)$, and 
$\alpha=0$. This particular choice
corresponds to a rotation about the $y$ axis. Let us change to the 
rotating basis,  $|\pm\rangle
\equiv (e^{-i\omega_B g^e_x t} e^{\dagger}_{+x}|G\rangle
\pm e^{i\omega_B g^e_x t} e^{\dagger}_{-x}|G\rangle)/\sqrt{2}$. The 
trion states are in the
rotating frame where $|T_i\rangle= e^{-i \omega t} 
|T_i^{(0)}\rangle$. The Hamiltonian in the basis
$|-\rangle,|+\rangle,|T_1\rangle,\cdots,|T_k\rangle$ becomes
\begin{equation} H=\left[\begin{array}{ccccc} 0 & 0 & 0 & 0 & 0 \\ 0 & 0 &
\Omega_1(t) &
\cdots & \Omega_k(t) \\ 0 & \Omega_1(t) & \Delta_1 & \cdots & 0 \\
\vdots & \vdots & \vdots & \ddots & \vdots \\ 0 & \Omega_k(t) & 0 &
\cdots & \Delta_k \\
\end{array}\right],
\end{equation} where $\Omega_k=\beta_k \Omega_{1}(t)/\sqrt{2}$, and
$\beta_k$ is the oscillator strength of the optical transition.   The 
first eigenvalue is zero
$\lambda_1(t)=0$. The second eigenvalue $\lambda_2(t)$ can be 
calculated exactly.  However, it
often suffices to work in the second order perturbation theory in 
which the analytic expression is
\begin{equation}
\lambda_2(t)= - \sum_i \frac{|\Omega_i(t)|^2}{\Delta_i}.
\end{equation} The corresponding eigenstates  are

\begin{eqnarray} |\lambda_1(t)\rangle &=&|-\rangle \\ |\lambda_2(t)\rangle
&=&|+\rangle-\sum_{i=1}^k  |T_i\rangle\frac{\Omega_i(t)}{\Delta_i} \\ 
|\lambda_{i+2}(t)\rangle
&=&|T_i\rangle+ |+\rangle
\frac{\Omega_i(t)}{\Delta_i}
\end{eqnarray} The adiabatic condition can be expressed as $|\langle
\lambda_i(t) |\frac{d}{dt}|\lambda_2(t)\rangle| \ll 
|\lambda_i(t)-\lambda_2(t)|$, for all  $i > 2$.
The most stringent condition is of course for the lowest trion state 
which gives
\begin{equation}
\frac{\dot{\Omega}_1(t)}{\Delta_1} \ll \Delta_1.
\label{admulti}
\end{equation} When this condition is fulfilled spin rotation can be 
achieved via multiple
intermediate trion states. The coherence of the rotation is again 
preserved by the virtual
excitation of intermediate states.

\subsection{Continuum $\Lambda$ system}
\label{raman_continuum}

STIRAP via continuum has been proposed for population transfer in atomic
physics.\cite{vitanov97,yatsenko97} Several approaches have been 
proposed to avoid leakage and
decoherence.\cite{decay} We show here how the adiabatic manipulation 
of a single spin can be
realized in principle in presence of a continuum. The key for 
avoiding leakage and decoherence is
again an excitation below the continuum edge.  The continuum is thus 
only virtually excited and the
coherence of the spin rotation is preserved.  The treatment parallels 
that of the multiple
$\Lambda$ system case  in section~\ref{raman_4level}.  The eigenenergy of
$|\lambda_1(t)\rangle=|-\rangle$ is again
$\lambda_1(t)=0$. By means of Fano's method\cite{fano61} the 
eigenenergy of the other discrete
state $\lambda_2(t)$ can be determined by the integral equation,
\begin{equation}
\lambda_2(t)=\int d\epsilon g(\epsilon)
\frac{|\Omega(\epsilon)|^2}{(\lambda_2(t)-\Delta_\epsilon)} ,
\end{equation} where we have replaced the summation over the discrete 
$k$ by the integral over the
energy with the density of states $g(\epsilon)$. At
$t=-\infty$ the state $|\lambda_2(-\infty)\rangle=|+\rangle$. The 
eigenvector of the new discrete
and continuum states $|\lambda_\epsilon(t)\rangle$ can be solved 
analytically (not shown here).
The adiabatic condition then can be expressed as
\begin{equation} |\langle
\lambda_\epsilon(t)|\frac{d}{dt}|\lambda_2(t)\rangle| \ll
\Delta_\epsilon~.
\label{adfano}
\end{equation}  The Fano approach thus allows us to obtain an 
analytical but complicated expression
for this condition valid to all orders in
$\Omega/\Delta$.  However, a more stringent condition can be obtained 
by expanding
Eq.~(\ref{adfano}) to second order in $\Omega/\Delta$ as is done for 
the multiple$\Lambda$ case.
It can be shown that it is sufficient to require
$\dot{\Omega}_{\epsilon=0}(t)/\Delta \ll \Delta$ where $\Delta$ is 
the detuning to the continuum
edge to fulfill the adiabatic condition, which is analogous to 
Eq.~(\ref{admulti}).  When this
condition is fulfilled it is possible to perform an adiabatic Raman 
transition coherently via the
continuum of intermediate states.

\section{Conclusions}
\label{conclusion}

We have presented a  theory for arbitrary rotations of the spin of a 
single electron in a
quantum dot via Raman transitions in the adiabatic limit.  Charged 
exciton states, or trions, play
the role of the upper level in an effective lambda system. An 
arbitrary spin rotation may be
performed by tailoring the relative phase and the relative 
intensities of two laser pulses as well
as choosing the polarization of the light and the orientation of a 
static magnetic field.  The
explicit relations between the parameters of the laser pulses and the 
angle and the axis of the
spin rotation are given.  We investigate how the intermediate state 
decoherence is suppressed when
the operations are performed in the adiabatic regime. We derive the 
adiabatic condition in lambda
systems where additional discrete levels or a continuum of states are 
present. We show the
calculations for a representative case (the monolayer fluctuation 
quantum dots) with values of the
oscillator strengths and the characteristic energy separation taken 
from the experiments.  We emphasize
that  our scheme works independently of the confinement properties of 
the dots, as long as the
structure of the trion levels can be represented by the one in 
Fig.~1a or Fig.~1b. In principle, quantum
dots could be engineered to optimize the fidelity of the operations 
with this particular control
scheme. The theory developed here provides a useful blueprint for the 
realization of single qubit
operations in spin-based quantum information processing.

\acknowledgments This work was supported by ARDA/ARO-DAAD19-01-1-0478,
  NSF DMR-0099572, and DARPA SpinS program. We thank Sophia Economou 
and Lucasz Cywinski for
careful readings of the draft.


\begin{thebibliography}{}

\bibitem{yafet73} Y. Yafet in {\it New developments in 
semiconductors}. Edited by P. R. Wallace, R.
Harris and M. J. Zuckermann. (Noordhoff, Leyden 1973).

\bibitem{thomas68} D. G. Thomas and J. J. Hopfield, Phys. Rev. {\bf 
175}, 1021 (1968).

\bibitem{hu76} P. Hu, S. Geschwind, and T. M. Jedju, Phys. Rev. Lett. 
{\bf 37}, 1357 (1976).

\bibitem{geschwind84} S. Geschwind and R. Romestain in {\it Light 
Scattering in Solids IV}. Edited
by M. Cardona and G. G\"untherodt. (Springer-Verlag, Berlin 1984).


\bibitem{ima} A. Imamoglu {\it et al.}, Phys. Rev. Lett. {\bf 83}, 4204 (1999).
\bibitem{plenio} D. Jonathan, M.B. Plenio and P.L. Knight, Phys. Rev. 
A {\bf 62}, 042307 (2000).
\bibitem{wine} D.J. Wineland {\it et al.}, arXiv:quant-ph/0212079 v2 (2003).
\bibitem{lidar} R. Zadoyan, D. Kohen, D.A. Lidar, and V.A. Apkarian, 
Chem. Phys. 266, 323 (2001).
\bibitem{ham} P.R. Hammer, A. V. Turukhin, M. S. Shahriar, and J. A. 
Musser, Opt. Lett. {\bf 26}, 361
(2001).
\bibitem{awsch} J.A. Gupta, R. Knobel, N. Samarth, D.D. Awschalom, 
Science {\bf 292}, 2459 (2002).
\bibitem{merlin} J. Bao A.V. Bragas, J.K. Furdyna and R. Merlin, 
Nature-Materials {\bf 2},
175-179 (2003).

\bibitem{chen01} P. Chen, C. Piermarocchi, and L. J. Sham, Phys. Rev. 
Lett. {\bf 87}, 067401 (2001).

\bibitem{pier02} C. Piermarocchi, P. Chen, Y.S. Dale, and L. J. Sham, 
Phys. Rev. B {\bf 65}, 075307
(2002).

\bibitem{stievater01} T. H. Stievater {\it et al.}, Phys. Rev. Lett. 
{\bf 87} 133603 (2001).

\bibitem{kamada01} H. Kamada {\it et al.}, Phys. Rev. Lett. {\bf 87} 
246401 (2001).

\bibitem{htoon02} H. Htoon {\it et al.}, Phys. Rev. Lett. {\bf 88} 
087401 (2002).

\bibitem{wojs} A. Wojs and P. Hawrylak, Phys. Rev. B {\bf 51}, 10880 (1995).

\bibitem{landin} L. Landin et al., Science {\bf 280}, 262 (1998).

\bibitem{hartmann} A. Hartmann et al., Phys. Rev. Lett. {\bf 84}, 5648 (2000).

\bibitem{tischler02} J. G. Tischler, A. S. Bracker, D. Gammon and D. 
Park, Phys. Rev. B {\bf 66},
081310(R) (2002).

\bibitem{loss98} D. Loss and D. P. Di Vincenzo, Phys. Rev. A {\bf 
57}, 120 (1998).

\bibitem{gupta99} J. A. Gupta, D. D. Awschalom, X. Peng, and A. P. 
Alivisatos, Phys. Rev. B {\bf
59}, R 10 421 (1999).

\bibitem{carlo02} C. Piermarocchi, P. C. Chen, L. J. Sham, and D. G. 
Steel, Phys. Rev. Lett {\bf
89} 167402 (2002).

\bibitem{divincenzo00} D. Bacon, J. Kempe, D.A. Lidar and K.B. 
Whaley, Phys. Rev. Lett. {\bf 85},
1758 (2000); D. P. Di Vincenzo, D. Bacon, J. Kempe, G. Burkard, K. B. 
Whaley, Nature {\bf 408}, 339
(2000).

\bibitem{pazy03} E. Pazy {\it et al.}, Europhys. Lett. {\bf 62}, 175 (2003).

\bibitem{gammon96} D. Gammon {\it et al.}, Science {\bf 273}, 87 
(1996); M. W. Bertz {\it et al.}
Solid State Commun. {\bf 86} 43 (1993).

\bibitem{yang} S.-R. Eric Yang and L.J. Sham, Phys. Rev. Lett. {\bf 
58}, 2598 (1987).

\bibitem{STIRAP} For review see K. Bergmann, H. Theuer, and B. W. 
Shore, Rev. Mod. Phys. {\bf 70},
1003 (1998); N. V. Vitanov, T. Halfmann, B. W. Shore, and K. 
Bergmann, Annu. Rev. Phys. Chem. {\bf
52}, 763 (2001).

\bibitem{vitanov97} N. V. Vitanov and S. Stenholm, Phys. Rev. A {\bf 
55}, 648 (1997).

\bibitem{unanyan01} R. G. Unanyan, B. W. Shore, and K. Bergmann, 
Phys. Rev. A {\bf 63}, 043405
(2001).

\bibitem{kis02} Z. Kis and F. Renzoni, Phys. Rev. A {\bf 65}, 032318 (2002).

\bibitem{troiani03} F. Troiani, E. Molinari and U. Hohenester, Phys. 
Rev Lett. {\bf 90}, 206802
(2003).

\bibitem{calarco03} T. Calarco, A. Datta, P. Fedichev, E. Pazy, and 
P. Zoller, Phys. Rev. A {\bf 68},
012310 (2003).

\bibitem{lindblad} G. Lindblad, Commun. Math. Phys. {\bf 48}, 119 (1976).

\bibitem{carmichael} H. Carmichael, {\it An open system approach to 
quantum optics},
(Springer-Verlag, Berlin 1993).

\bibitem{bonadeo} N.H. Bonadeo {\em et al.}, Phys. Rev. Lett. {\bf 
81}, 2759 (1998).

\bibitem{vitanov99} N. V. Vitanov and S. Stenholm, Phys. Rev. A {\bf 
60}, 3820 (1999).

\bibitem{yatsenko97} L. P. Yatsenko, R. G. Unanyan, K. Bergmann, T. 
Halfmann, and B. W. Shore, Opt.
Commun. {\bf 135}, 406 (1997).


\bibitem{decay} R. G. Unanyan, N. V. Vitanov, B. W. Shore, K. 
Bergmann, Phys. Rev. A {\bf 61}
043408 (2000); R. G. Unanyan, N. V. Vitanov, and S. Stenholm, Phys. 
Rev. A {\bf 57} 462 (1998).


\bibitem{fano61} U. Fano, Phys. Rev.  {\bf 124}, 1866 (1961).

\end{thebibliography}
\end{document}